# Distinct Composition-Dependent Topological Hall Effect in Mn$_{2-x}$Zn$_x$Sb

Md Rafique Un Nabi[1,2], Yue Li[3], Suzanne G E te Velthuis[3], Santosh Karki Chhetri[1], Dinesh Upreti[1], Rabindra Basnet[1,5], Gokul Acharya[1], Charudatta Phatak[3,4], Jin Hu[1,2,6*]

[1]Department of Physics, University of Arkansas, Fayetteville, AR 72701, USA

[2]MonArk NSF Quantum Foundry, University of Arkansas, Fayetteville, Arkansas 72701, USA

[3]Materials Science Division, Argonne National Laboratory, Lemont, Illinois 60439, USA

[4]Department of Materials Science and Engineering, Northwestern University, Evanston, Illinois 60208, USA

[5]Department of Chemistry & Physics, University of Arkansas at Pine Bluff, Pine Bluff, Arkansas 71603, USA

[6]Materials Science and Engineering Program, Institute for Nanoscience and Engineering, University of Arkansas, Fayetteville, AR 72701, USA

## Abstract

Spintronics, an evolving interdisciplinary field at the intersection of magnetism and electronics, explores innovative applications of electron charge and spin properties for advanced electronic devices. The topological Hall effect, a key component in spintronics, has gained significance due to emerging theories surrounding noncoplanar chiral spin textures. This study focuses on Mn$_{2-x}$Zn$_x$Sb, a material crystallizing in centrosymmetric space group with rich magnetic phases tunable by Zn contents. Through comprehensive magnetic and transport characterizations, we found that the high-Zn ($x > 0.6$) samples display THE which is enhanced with decreasing temperature, while THE in the low-Zn ($x < 0.6$) samples show an opposite trend. The coexistence of those distinct temperature dependences for THE suggests very different magnetic interactions/structure for different compositions and underscores the strong coupling between magnetism and transport in $Mn_{2-x}Zn_xSb$. Our findings contribute to understanding topological magnetism in centrosymmetric tetragonal lattices, establishing $Mn_{2-x}Zn_xSb$ as a unique platform for exploring tunable transport effects and opening avenues for further exploration in the realm of spintronics.

## 1. Introduction

Spintronics, an interdisciplinary field at the nexus of magnetism and electronics, continues to undergo rapid development. It delves into the innovative use of electron charge and spin properties for advanced electronic devices. A key component, the topological Hall Effect (THE), is increasingly recognized for its significance in the realm of spintronics, propelled by emerging theories surrounding noncoplanar chiral spin textures. These chiral spin textures are attributed as the primary mechanism driving THE, and their existence is related to the Berry-phase accumulation in real space [1,2]. When itinerant electrons hop over a chiral spin texture, they generate an effective magnetic flux due to the real space Berry phase [3,4]. This fictitious magnetic field manifests as an additional transverse voltage drop known as the THE. The THE is considered as a distinctive hallmark of materials possessing topologically protected spin structures in real space, where the spin texture induces an effective magnetic field that influences the trajectories of itinerant electrons. These exotic electromagnetic properties have sparked great interests both for theoretical research on magnetic monopoles [5] and gauge field theory [6] in condensed matter physics and experimental investigation on the potential information carriers in the next generation storage and logic devices [7–9].

THE and the existence of chiral spin textures have been widely studied in bulk non-centrosymmetric lattice structures, like (Mn, Fe, Co)(Si, Ge) [10–24], $Cu_2OSeO_3$ [25,26], $\beta$-Mn-type Co-Zn-Mn [27,28], and polar $GaV_4S_8$ [29] and $VOSe_2O_5$ [30]. Most of materials exhibiting THE are ferromagnetic semiconductors with non-centrosymmetric B20 structures, also known as chiral magnets, governed by Dzyaloshinskii-Moriya interactions induced by spin-orbit coupling. Interestingly, the THE has also been identified in a few centrosymmetric magnetic materials. Recent studies have found that some centrosymmetric noncollinear magnets display THE as well, such as the complex noncollinear ferromagnet $NdMn_2Ge_2$ [31], canted ferromagnet MnNiGa [32], the frustrated kagome ferromagnet $Fe_3Sn_2$ [33], and frustrated triangular antiferromagnet $Gd_2PdSi_3$ [34]. THE in those material is reported to arise from skyrmion lattices. Unlike non-centrosymmetric materials, skyrmions are stabilized in those centrosymmetric materials by the interplay of the external magnetic field, exchange interaction, uniaxial magnetic anisotropy, magnetic dipole, and Ruderman-Kittle-Kasuya-Yoswida-type interactions [31,35–37].

Here we focus on THE in tetragonal system $Mn_{2-x}Zn_xSb$. $Mn_{2-x}Zn_xSb$ can be derived from its parent compound $Mn_2Sb$, which is a well-known material crystallized in a centrosymmetric layered tetragonal structure with a space group P4/*nmm*. $Mn_2Sb$ comprises two magnetic sublattices Mn(I) and Mn(II). Mn moments within each sublattice are parallel to each other, but the moments for Mn(I) and Mn(II) sublattice are antiparallel and non-equal, leading to ferrimagnetic (FIM) order. Below 550K, $Mn_2Sb$ exhibits ferrimagnetic order with Mn(I) and Mn(II) moments aligned along the *c*-axis. Cooling to 240K, a spin-reorientation occurs, which is characterized by the rotation of the moments of both Mn(I) and Mn(II) sublattices along the *ab*-plane. Such magnetic transition keeps $Mn_2Sb$ ferrimagnetic [38]. The magnetism in $Mn_2Sb$ is tunable upon metal substitution. Transition metal elements such as Zn [39–43], Cr [44–49], Co [50–54], or Fe [55,56] have been found to replace the Mn(II) sublattice. Among various substitutions, the nonmagnetic Zn substitution creates $Mn_{2-x}Zn_xSb$ which is particularly interesting. Such substitution does not change the centrosymmetric lattice, and the full replacement of Mn(II) by Zn leaves only the ferromagnetic Mn(I) magnetic lattice, leading to room temperature ferromagnetism in MnZnSb [41,57]. Moreover, the layered structure of $Mn_{2-x}Zn_xSb$ facilitate exfoliation to thin flakes, which potentially provides a new member in low-dimensional magnetism and offers an additional candidate for heterostructure integration to take advantage of high magnetic ordering temperature. Compared to MnZnSb in which the Mn(II) sublattice is fully substituted by Zn, partially Zn-substituted $Mn_{2-x}Zn_xSb$ compounds exhibit rich magnetic phases and unusual transport properties [39,40,42,43]. In particular, our previous study has demonstrated giant THE in high-Zn content $Mn_{2-x}Zn_xSb$ samples, implying the presence of non-trivial spin textures [42]. Given that the high-Zn ($x > 0.6$) and low-Zn ($x < 0.6$) samples exhibit distinct transport properties characterized by metal-to-insulator transition-like and nearly fully metallic behaviors, respectively [43], it is rather interesting to extend the THE study to low-Zn samples.

Under this motivation, we investigated THE in high-Zn ($x > 0.6$) and low-Zn ($x < 0.6$) Mn$_{2-x}$Zn$_x$Sb samples in this work. Through comprehensive magnetic and transport property characterization, we found that while both groups of samples display THE, they display very different temperature dependence. THE is enhanced with rising temperature in low-Zn samples, but suppressed in high-Zn samples. Such difference might be attributed to the variation of magnetic interactions with Zn content, which is further confirmed by magneto-optic Kerr effect (MOKE) microscopy. These observations, together with the layered structure, establish Mn$_{2-x}$Zn$_x$Sb as a distinct material platform for tunable topological magnetism in centrosymmetric tetragonal lattices, offering further insight into 2D magnetism for spintronic applications.

## 2. Experimental Method

The single crystal growth for various Mn$_{2-x}$Zn$_x$Sb compositions has been reported in our previous work [43]. The crystal structure and composition were checked by using x-ray diffraction and energy dispersive x-ray spectroscopy (EDS), respectively. The EDS-determined compositions are used in the following discussions. The electronic transport measurements were conducted using a Physical Properties Measurement System (PPMS), and the magnetic properties were characterized by using a PPMS and a Magnetic Properties Measurement System (MPMS). Magnetic domains were imaged using a wide-field magneto-optical microscope from evico magnetics, optimized for the polar MOKE which is sensitive to the out-of-plane magnetization. Magnetic fields were applied using an electromagnet perpendicular to the surface being imaged. To maximize the magnetic contrast, an image taken at the electromagnet's maximum field (135 mT) was subtracted from the raw images. This maximum field was above or close to the magnetic saturation field of the samples.

3. Results and Discussions

As shown in Fig. 1, Mn$_2$Sb exhibits a tetragonal crystal structure with two inequivalent magnetic sublattices Mn(I) and Mn(II). Neutron studies have previously established that below 550K, these inequivalent Mn(I) and Mn(II) moments align antiparallel along the c axis, resulting a long range ferrimagnetic ordering [38]. As temperature further decreases below 240K, these Mn moments reorient along the *ab* plane to form a low temperature ferrimagnetic phase. In Mn$_{2-x}$Zn$_x$Sb, non-magnetic Zn substitutes at the Mn(II) site and maintains the *P*4/*nmm* space group and suppresses the magnetic transition to lower temperatures [42,43]. For low-Zn samples, an additional magnetic phase classified as weak ferrimagnetic phase occurs at lower temperature below the spin-reorientation transition temperature [39,43]. Such rich magnetic phases in Mn$_{2-x}$Zn$_x$Sb suggest the potential THE in this material, providing the emergence of chiral magnetic structures. Indeed, a giant topological hall effect with topological resistivity reaching $\sim 2$ $\mu\Omega$ cm in the $x \approx 0.85$ sample has been discovered [42]. This composition falls in the high-Zn composition region with $x > 0.6$, in which the samples display a metal-to-insulator-like transition at the magnetic transition temperature [43]. In contrast, the transport properties for low-Zn ($x < 0.6$) samples are drastically different, characterized by an overall metallic temperature dependence [43]. Therefore, whether or not THE occurs in low-Zn samples and how it differs from that in high Zn samples become natural questions.

We have performed Hall effect measurements for Mn$_{2-x}$Zn$_x$Sb with various Zn content covering the entire composition range. In Figs. 2a and 2b we show the Hall resistivity $\rho_{xy}$ measured at various temperatures from 2 to 350 K for two representative low-Zn samples $x = 0.3$ and $x = 0.48$. The results for representative high-Zn samples has been reported in our previous work [42]. The inevitable longitudinal component due to the asymmetric electrical contacts was removed by symmetrization using the measured data in positive and negative magnetic fields. These measurements clearly reveal strong anomalous Hall effect (AHE) in all $Mn_{2-x}Zn_xSb$ ($0 < x < 1$) samples, indicating strong coupling between magnetism and electronic transport, as manifested in low-field jumps in $\rho_{xy}$. At the resistivity jump, a peak-like feature signaturing THE indeed appears in those low-Zn samples, as indicated by the black arrows in Figs. 2a-b. Besides the AHE, our transport study also shows an additional component of Hall resistivity known as THE. Such THE arises when the moments in magnetic materials do not align in one direction but vary in position space. Therefore, the total Hall resistivity ($\rho_H$) including ordinary Hall effect (OHE), AHE and THE contributions can be expressed as

$$\rho_{xy} = R_0 H + S_H \rho_{xx}(H,T)^2 M(H,T) + \rho_{xy}^T \qquad (1),$$

where the first, second, and third terms correspond to the contributions from OHE, AHE and THE, respectively; $R_0$, $S_H$, $\rho_{xx}$, and $M$ represent ordinary Hall coefficient, anomalous Hall coefficient, longitudinal resistivity, and magnetization, respectively. Since $\rho_{xx}$ and $M$ can be experimentally measured, fitting the measured Hall resistivity $\rho_{xy}$ to the equation 1 can separate various Hall components. The field dependent $\rho_{xx}$ and $M$ measured at various temperatures are shown in Figs. 2e-c and Figs. 2e-f, respectively. With those, figure 3a shows one typical example for fitting the measured total Hall resistivity measured at 350 K for the $x = 0.30$ sample, from which the THE component (blue line in Fig. 3a) is separated from the total measured Hall resistivity. Using this approach, we have extracted THE components at different temperatures for $Mn_{2-x}Zn_xSb$ with various Zn contents. The results for the representative $x = 0.3$ and $x = 0.48$ samples are shown in Figs. 3c and 3d, respectively. In equation 1, the AHE component has a quadratic relation of longitudinal resistivity [i.e., $AHE \propto \rho_{xx}(B,T)^2$], which is applicable to extrinsic AHE in good metals or AHE with an intrinsic Berry phase mechanism. For the case of extrinsic AHE, such as skew scattering, a linear dependence (AHE $\propto \rho_{xx}$) should be adopted [18,58], which, however cannot reproduce the measured Hall resistivity.

From the extracted THE components shown in Figs. 3b and 3c for the representative $x = 0.3$ and $x = 0.48$ samples, THE occurs near zero field, which is consistent with the magnetization saturation at low fields (Figs. 2e and 2f). Such saturation indicates moment polarization that suppresses magnetic textures, leading $\rho_{xy}^T$ to vanish beyond the saturation field as observed (Figs. 3b and 3c). $\rho_{xy}^T$ in the $x = 0.48$ sample reaches ~0.9 µΩ cm at room temperature, which is roughly 3 times larger than that in the $x = 0.3$ sample at the same temperature. Such difference should correspond to the evolution of magnetism and spin texture with Zn content, as will be discussed later. Furthermore, $\rho_{xy}^T$ in both samples display an overall trend of suppression with lowering temperature, as shown in Figs. 3b and 3c.

For $x > 0.6$ high-Zn samples, our previous work [42] has found giant $\rho_{xy}^T$ ~2 µΩ cm for samples near $x = 0.85$. However, we found that such THE is fundamentally different from that in low-Zn samples. The large THE in high-Zn samples occurs at the lowest measured temperature ($T = 2$ K), which reduces with rising temperature and vanishes above 250 K. In Figs. 4a and 4b we have summarized the temperature dependencies for THE for low-Zn ($x > 0.6$) and high-Zn ($x > 0.6$) samples, respectively, based on which a contour plot is constructed, as shown in Fig. 4c. Distinct THE for those two groups of samples is clearly visualized. As shown in Fig. 4c, Zn ($x > 0.6$) and high-Zn ($x > 0.6$) samples display strong composition dependence for THE, and exhibit two distinct characteristics: First, THE in high-Zn samples reaches maximum for $x = 0.85$, and drops quickly when deviating from this composition. In contrast, remarkable THE exists in a much wider composition range in low-Zn samples. Second, THE suppresses with heating in high-Zn samples but enhances in low-Zn samples.

Both enhancement and suppression of THE with heating have been reported. For example, THE in FeGe [59], Mn$_2$NiGa [60], MnPdGa [61], Cr$_{1.2}$Te$_2$ [62], Fe$_{1-x}$Co$_x$Ge [23], La$_x$Eu$_{1-x}$O [63] is enhanced upon heating, which has been ascribed to chiral spin textures or skyrmion spin textures. Conversely, materials like Mn$_5$Si$_3$ [64], Gd$_2$PdSi$_3$ [34], and Mn$_2$RhSn [65] exhibit a suppression of THE with heating wherein a similar spin texture-based mechanism is implicated in the presence of THE. Those results indicate that the tempeature dependences for THE should be associated with the temperature dependence of spin texture. Though different temperature dependences have been reported in various materials, the coexistence of the opposite temperature dependences in the same material system has not been discovered. Therefore, Mn$_{2-x}$Zn$_x$Sb could be the first material system that displays such interesting phenomena, suggesting complicated magnetism that governing THE in low-Zn and high-Zn Mn$_{2-x}$Zn$_x$Sb samples. In fact, the critical composition of $x = 0.6$ has also been observed in our previous work [43]. At such composition the electronic transport properties modify drastically, suggesting a phase boundary that separate different magnetic and electronic phases [43]. In this work, the two distinct temperature dependencies for THE suggest potential differences in the underlying origins of THE within the two distinct composition groups.

Generally, THE occurs in the presence of spin textures or non-trivial band topology. A deep understanding of the distinct THE in low- and high-Zn samples require comprehensive information on magnetic structure and spin textures of those materials. Unfortunately, the magnetic structures are only experimentally determined for very limited compositions $x < 0.4$ [38,39]. Future neutron scattering expanding to Zn-rich samples will be very helpful. In this study, owing to the strong magnetism of $Mn_{2-x}Zn_xSb$, we have imaged the magnetic domains using polar MOKE microscopy measurements on single crystals. MOKE images taken at various temperatures for $x = 0.85$ and 0.3 single crystals, the representative high-Zn and low-Zn compositions, are shown in Figs. 5 and 6, respectively. As shown in Fig. 5, magnetic domains in the $x = 0.85$ sample are clearly visible even at zero or almost zero magnetic field at various temperatures up to 300 K, consistent with high magnetic ordering temperature above room temperature for this composition [42]. An evolution of the domain-type with temperature is observed. Labyrinth magnetic domains emerge at 300 K (Fig. 5d) and evolve into stripe-like magnetic domains but with a wider domain width when cooling down to 80 K (Fig. 5a). Interestingly, the crossover of the domain-type appears to occur at 250 K (Fig. 5c). Such temperature coincides with the temperature that THE start to develop (Fig. 4c). This suggests that a chiral spin texture may exist in these domains that leads to THE. In low-Zn samples, however, the magnetic domain configuration is completely different. For an $x = 0.30$ sample, there is not sufficient magnetic contrast to reveal a domain structure below 200 K. Above 200 K, the magnetic signal is still relatively weak and domains have relatively large and irregular shapes, as shown in Fig. 6, that is quite different from the bubble chain-like and labyrinth domains seen for the $x = 0.85$ sample. Nevertheless, the emergence of resolvable magnetic domains at high temperatures again coincides with the enhanced THE upon rising temperature in low-Zn samples. While the images shown in Figs. 5 and 6 are representative for samples with high and low Zn doping, we did find some variations in the domain structure for different samples with nominally the same value of x, implying composition dependent magnetism in $Mn_{2-x}Zn_xSb$.

The consistent temperature dependencies for THE and magnetic domains in high-Zn and low-Zn samples suggest the role of chiral spin texture in the domain or domain wall in mediating THE. It is also highly interesting to find chiral magnetic structure or skrymion lattice in $Mn_{2-x}Zn_xSb$ which crystalizes in a centrosymmetric tetragonal lattice. Nevertheless, It is worth nothing that $Mn_{2-x}Zn_xSb$ is formed by the substitution of the Mn(II) magnetic sublattice with Zn in $Mn_2Sb$ (see Fig. 1). Given that $Mn_2Sb$ is ferrimagnetic, the low-Zn samples are essentially Mn(II)-rich samples, in which the magnetization is reduced owing to the opposite magnetic moment orientations for Mn(I) and Mn(II). The MOKE measurements reveal difference in magnetic domain structures in low-Zn and high-Zn samples, implying different magnetism that may give rise to the distinct temperature dependence of THE. Therefore, understanding the underlying mechanism behind distinct magnetism is crucial. Providing $Mn_{2-x}Zn_xSb$ is a chemically substitute system with rich magnetic phases, magnetic and structural disorders might contribute to the formation of different spin textures. The substitution of non-magnetic Zn for Mn(I) disturbs the magnetism, including magnetic disorders in $Mn_{2-x}Zn_xSb$ that is manifested into moment canting in local regions. The possible variation of the canting angle and canting region size with temperature may lead to temperature dependent THE. In addition, the differences in atomic size and electronegativity between Zn and Mn could result in local structural modifications, which may lead to slight atom displacements and contributing to the emergence of DMI, which further results in non-trivial spin texture. To find the actual mechanism behind the distinct THE in low- and high-Zn samples, additional experimental studies are needed to clarify the structure and magnetism of this material system, such as high-resolution transmission electron microscopy, Lorentz transmission electron microscopy, and neutron scattering, as well as theoretical efforts.

4. Conclusion

In summary, our study on $Mn_{2-x}Zn_xSb$ reveals composition-dependent THE. The high-Zn samples exhibit an increase in THE with decreasing temperature, while the low-Zn samples show an opposite trend. This dichotomy features the influence of Zn substitution on the magnetic phases and transport properties in $Mn_{2-x}Zn_xSb$. The observed temperature variations in THE, coupled with the existence of non-trivial spin textures, highlight the strong coupling between magnetism and transport in this material. The layered structure of $Mn_{2-x}Zn_xSb$ also offers opportunities for microexfoliation and thin film growth, enabling the implementation of those intriguing transport and magnetic properties in device applications. Our findings contribute to the exploration of topological magnetism in centrosymmetric tetragonal lattices, establishing $Mn_{2-x}Zn_xSb$ as a good platform for exploring tunable topological effects.

**Acknowledgements**
This work was supported by the MonArk NSF Quantum Foundry supported by the National Science Foundation Q-AMASE-i program under NSF award No. DMR-1906383. Work at the Argonne National laboratory (MOKE microscopy) was supported by the US Department of

Energy, Office of Science, Office of Basic Energy Sciences, Materials Science and Engineering Division. J.H. acknowledges the support from NSF under Grant No. DMR-2238254.

Figure 1

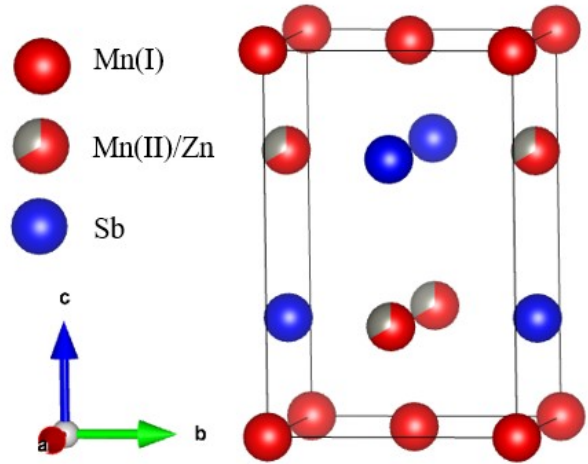

FIG. 1. (a) Crystal Structure of $Mn_{2-x}Zn_xSb$.

Figure 2

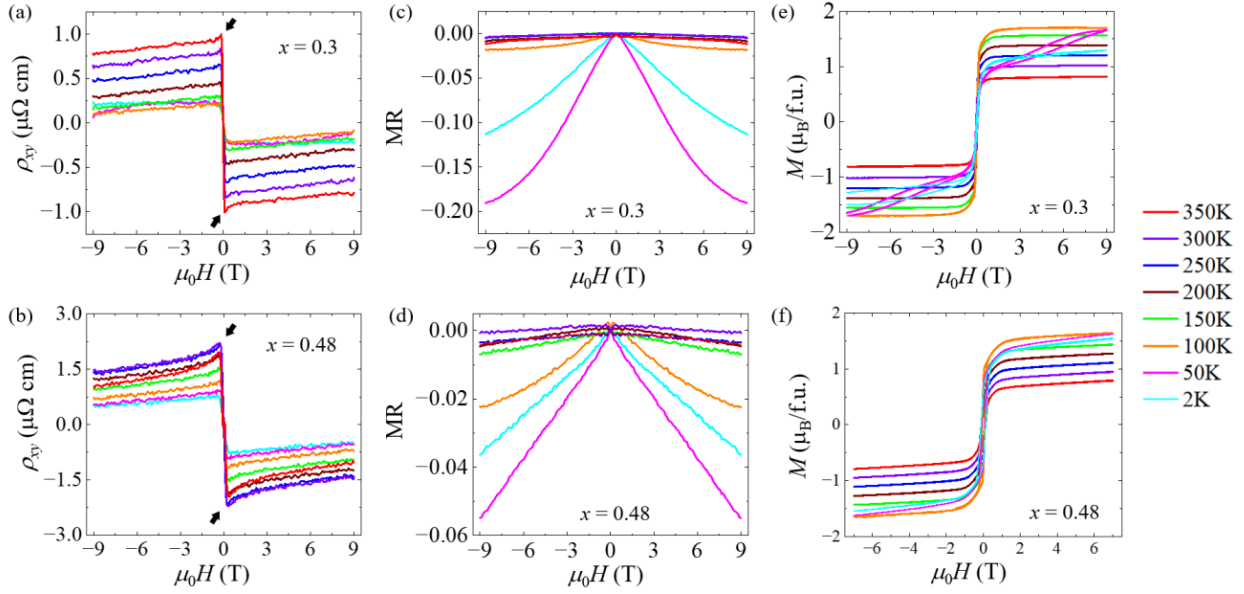

FIG. 2. Magnetic field dependence at various temperatures for $x = 0.3$ and $x = 0.48$ samples for: (a) and (b), isothermal Hall resistivity $\rho_{xy}$, the black arrows indicates the emergence of THE; (c) and (d), normalized magnetoresistivity, MR = $[\rho_{xx}(H) - \rho_{xx}(H=0)/\rho_{xx}(H=0)]$; and (e) and (f), out-of-plane ($H \parallel c$) magnetization $M$.

Figure 3

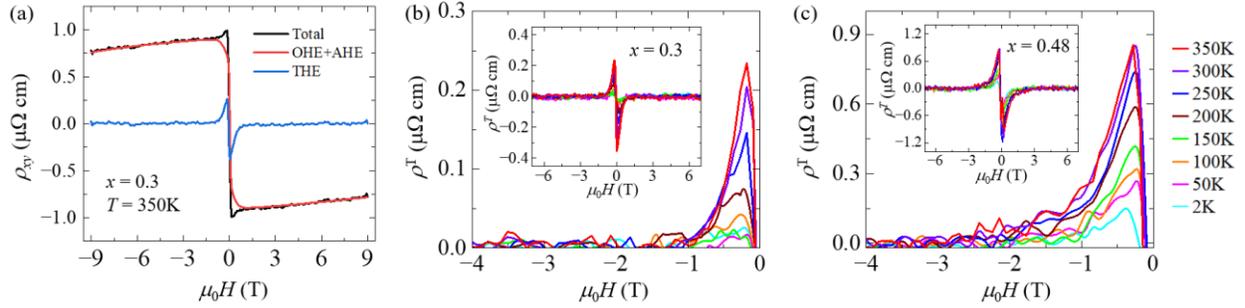

FIG. 3: (a) Separation of ordinary (OHE), anomalous (AHE) and topological (THE) Hall components for the $x = 0.30$ sample at $T = 350$K. (b) and (c) Extracted THE for $x = 0.30$ and $x = 0.48$ samples, respectively, at different temperatures. The insets show the entire field range, and the main panels show the zoom in near the zero field where THE occurs.

Figure 4

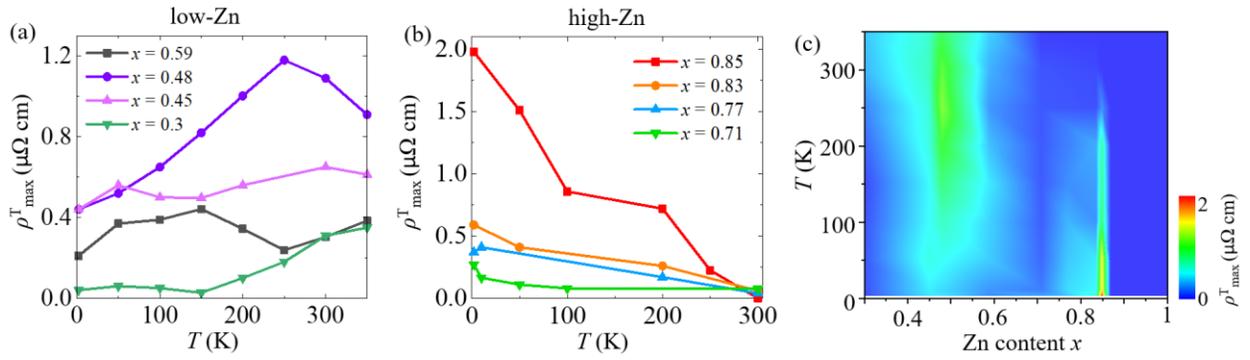

FIG. 4: (a) and (b) Maximum topological Hall resistivity for low-Zn ($x < 0.6$) and high-Zn ($x > 0.6$) samples at various temperatures; (c) Contour map showing the temperature and composition dependence for the maximum topological hall resistivity for $Mn_{2-x}Zn_xSb$. Data for Zn-rich samples from Ref. 54 are included to show the complete composition and temperature dependence.

Figure 5

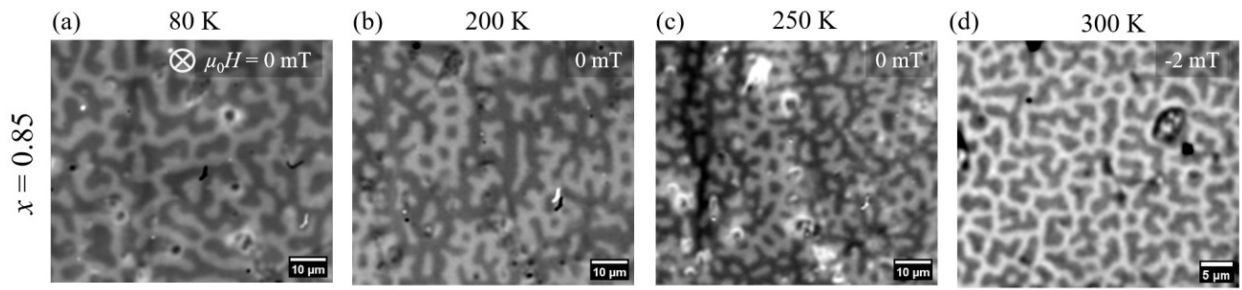

FIG. 5: Polar MOKE imaging of magnetic spin textures for a $x = 0.85$ single crystal at different temperatures. The orientation of magnetic field is perpendicular to the sample surface.

Figure 6

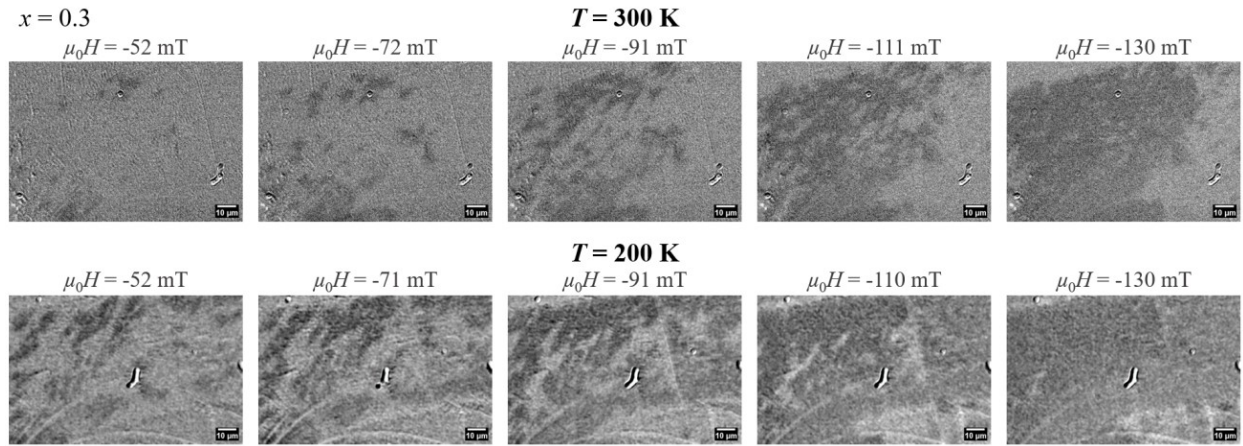

FIG. 6: Polar MOKE imaging of magnetic spin textures for a $x = 0.3$ single crystal at different magnetic fields at 300 K (top) and 200 K (bottom). The orientation of magnetic field is perpendicular to the sample surface.